\documentclass[a4paper,11pt,fleqn]{article}
\usepackage[left=2.5cm,right=2.5cm,top=2cm,bottom=2.5cm]{geometry} 
\usepackage[bf,font={singlespacing,small}]{caption}
\usepackage{setspace,placeins,cite}

\usepackage[affil-it]{authblk}
\usepackage[italian]{babel}
\usepackage[latin1]{inputenc}
\usepackage{epsfig,graphics,graphicx,color}
\usepackage{amsmath,amssymb,amsthm,latexsym,mathrsfs}

\usepackage[bookmarks=false, %
            pdftitle={Father Eliseo della Concezione}, %
            pdfauthor={Aurelio Agliolo Gallitto}, %
            pdftex, %
            hyperfootnotes=false, %
            pdfpagelabels, %
            pdfstartview=FitH, %
            hidelinks]{hyperref}

\vspace{-2.5cm}

\begin{document}

\title{\vspace{-1.5cm}\textbf{$\mathbf{300^{th}}$ anniversary of the birth of Father Eliseo della Concezione: the scientific contribution in the Royal Academy of Studies of Palermo}}

\author{Aurelio Agliolo Gallitto}

\affil{Dipartimento di Fisica e Chimica - Emilio Segrè, University of Palermo, Palermo, Italy, e-mail:~\href{mailto:aurelio.agliologallitto@unipa.it}{aurelio.agliologallitto@unipa.it}}

\date{}

\maketitle

\vfill

\begin{spacing}{1.2}

\noindent \begin{center}\textbf{\centering \Large Abstract} \end{center} \smallskip

\noindent 2025 marks the $\mathrm{300^{th}}$ birthday of Father Eliseo della Concezione, professor of Experimental Physics at the Royal Academy of Studies of Palermo.
To celebrate this anniversary, the Physics and Chemistry Library of the University Library System and the Department of Physics and Chemistry - Emilio Segrè have organized several cultural activities.
In the article, after a brief biographical description of Father Eliseo della Concezione, we will present the activities carried out and discuss the historical and educational aspects of Father Eliseo's work carried out during his stay at the Royal Academy of Palermo at the end of $\mathrm{18^{th}}$ century.

\bigskip


\vfill

\noindent \begin{center} \textbf{\centering \Large \emph{Riassunto}} \end{center} \smallskip

\noindent \emph{300° anniversario della nascita di padre Eliseo della Concezione: il contributo scientifico nella Regia Accademia degli Studi di Palermo}
\bigskip

\noindent \emph{Nel 2025 ricorre il 300° anniversario della nascita di padre Eliseo della Concezione, docente di Fisica Sperimentale nella Regia Accademia degli Studi di Palermo.
Per celebrare questo evento, la Biblioteca di Fisica e Chimica del Sistema Bibliotecario e Archivio Storico di Ateneo e il Dipartimento di Fisica e Chimica - Emilio Segrè hanno organizzato varie attività culturali.
Nell'articolo, dopo una breve descrizione biografica di Padre Eliseo della Concezione, presenteremo le attività svolte e discuteremo gli aspetti storico-didattici del lavoro svolto da padre Eliseo nella Regia Accademia di Palermo alla fine del XVIII secolo.}

\bigskip

\noindent \emph{\textbf{Parole chiave:} Storia della Fisica, Eliseo della Concezione, Patrimonio Scientifico e Tecnologico, Hands-on Science}

\end{spacing}

\vfill

\newpage

\section{Introduzione}
Tra le attività di terza missione condotte dall'Università degli Studi di Palermo, vi è la ``Settimana delle biblioteche'', organizzata annualmente dal Sistema Bibliotecario e Archivio Storico di Ateneo (SBA) dal 2018.
Una settimana dedicata alla promozione e valorizzazione del patrimonio documentario custodito nelle biblioteche dell'Ateneo palermitano, attraverso mostre, esposizioni, seminari, incontri di lettura, ecc.
L'evento, aperto alla cittadinanza, si inserisce nel programma nazionale ``Il Maggio dei Libri'', promosso dal Centro per il Libro e la Lettura del Ministero della Cultura.
In queste occasioni, per coinvolgere attivamente i visitatori, per la maggior parte studenti delle scuole secondarie del territorio, vengono organizzate esposizioni tematiche, visite guidate alle collezioni universitarie e laboratori didattici.
Nell'edizione che si è svolta dal 5 al 9 maggio 2025, la Biblioteca di Fisica e Chimica in collaborazione con il Dipartimento di Fisica e Chimica - Emilio Segrè (DiFC) hanno organizzato varie attività culturali per la ricorrenza del 300° anniversario della nascita di padre Eliseo della Concezione, nato a Napoli il 16 agosto 1725~\cite{Pagano1993}, docente di Fisica Sperimentale nella Regia Accademia degli Studi di Palermo dal 1786 al 1793~\cite{Nastasi1998}.
È stata allestita un'esposizione bibliografica e archivistica riguardante p.~Eliseo, organizzate visite guidate alla Collezione Storica degli Strumenti di Fisica e laboratori didattici ``Hands-on Science''.
Classi di scuola secondaria sono state invitate a partecipare all'evento, dando la possibilità agli studenti, sotto la guida dei loro insegnanti, di svolgere i laboratori didattici e sperimentare con gli exhibit allestiti per l'occasione.

\section{Introduzione storica}
Dopo l'istituzione della \emph{Deputazione de' Regii Studj}, l'1 agosto 1778, alla quale Ferdinando di Borbone (1759~-~1816), re di Sicilia, aveva assegnato il compito di riordinare il sistema scolastico siciliano, l'1 maggio 1779, con decreto reale, fu istituita la \emph{Regia Accademia degli Studi di Palermo}~\cite{Cancila2006,Zingales2022}, nella sede dell'ex Collegio Massimo dei Gesuiti al Cassaro \mbox{--~oggi} biblioteca centrale della Regione Siciliana.
L'Accademia era composta da 4 facoltà e 20 cattedre, tra cui la cattedra di Fisica Sperimentale con annesso Gabinetto per effettuare gli esperimenti~\cite{Nastasi1998}.
In seguito, il 12 gennaio 1806 fu fondata l'\emph{Università degli Studi}~\cite{web1}.

In un piano delle cattedre attive nell'ottobre 1778, nell'ex Collegio Massimo dei Gesuiti, risultava che la funzione di lettore di Fisica venisse esercitata interinamente da un certo Don Nicolò Fresco, medico napoletano~\cite{Cancila2006,Romano2006}.
Tuttavia, non godendo il Fresco di una buona reputazione, per le sue frequenti e prolungate assenze, nell'anno accademico 1779/80, la Deputazione incaricò come lettore interinario di Fisica Sperimentale il padre domenicano Antonio Minasi O.~P.~(1736~-~1806), il quale fu nuovamente sostituito dal Fresco, probabilmente per indisponibilità di p.~Minasi a tenere l'incarico, considerato l'esiguo compenso assegnato alla cattedra~\cite{Cancila2006}.

Grazie all'interessamento del principe di Cimitile, ambasciatore napoletano a Roma, la Deputazione riuscì ad assicurarsi la disponibilità del padre napoletano Eliseo della Concezione (1725~-~1809), nominato nel luglio 1780, con un salario annuo di 240 ducati (80 onze).
In precedenza, la Deputazione aveva ricevuto ``\emph{favorevoli rapporti}'' su p.~Eliseo:

\begin{quote}
``\emph{...reputato uno dei più intesi della facoltà fisica sperimentale e ben pratico \mbox{nell'esercizio} delle macchine e degli esperimenti.}''
\end{quote}

\noindent Tuttavia, p.~Eliseo continuava a trattenersi a Napoli, cosicché nel febbraio 1781 la Deputazione decise di sostituirlo con padre Salvatore da Santa Maria, dell'ordine degli Scalzi della Mercede, ``\emph{della cui letteratura e abilità nella fisica sperimentale ne abbiamo avuti ottimi rapporti}'', con un salario di 60 onze l'anno~\cite{Cancila2006,Romano2006}.
A conferma di ciò, in un ``\emph{Regiae Panormitanae UNIVERSITATIS anni MDCCLXXXI Studiorum Conspectus}'' non si trova traccia di lezioni tenute dal Fresco, si trovano invece indicate le lezioni, tenute in ``\emph{Hora Vespertina}'', di p.~Salvatore da Santa Maria~\cite{Nastasi1998}:

\begin{quote}
``\emph{Physicae Institutiones perpetuis a Georgio Atwood Experimentis confirmatas tradet, atque illustrabit.}''
\end{quote}

\noindent L'abate Domenico Scinà (1764-1837), autore del \emph{Prospetto della storia letteraria di Sicilia}~\cite{Scina1859}, precisa che p.~Salvatore~\cite[p.~355]{Scina1859}

\begin{quote}
``\emph{leggeva la fisica non senza qualche decoro, e alcune principali esperienze nelle sue lezioni recava con l'assistenza di Giovanni Francone.}''
\end{quote}

\noindent Nel citato \emph{Conspectus} egli è, infatti, indicato quale tecnico di laboratorio:

\begin{quote}
``\emph{Physicis, ac Mathematicis Experimentis inserviet Instrumentorum in Academia Machinator D. JOANNES FRANCONE.}''
\end{quote}

\noindent Originario della Lombardia, Francone era noto quale esperto costruttore di strumenti di fisica (barometri e termometri e qualche altra macchina).

Seppure vada a essi attribuito il merito di aver diffuso le conoscenze di base della fisica sperimentale, Scinà commenta~ \cite[pp.~355-356]{Scina1859}:

\begin{quote}
``\emph{Comuni erano tra noi gli esperimenti così della macchina pneumatica, come dell'elettrica...}'',
\end{quote}

\noindent tuttavia nei primi anni dell'Accademia

\begin{quote}
``\emph{...si studiava la fisica più colla teorica che colle macchine, né questa di altro occupavasi, che delle esperienze principali, che si operavano colla macchina elettrica e pneumatica.}''
\end{quote}

\noindent Come dice ancora Scinà, in Sicilia la ricerca nel campo della fisica si era arrestata a vecchie concezioni di origine scolastica; mancavano totalmente sia metodo sia sperimentazione.

\begin{figure}[h!]
\centering
\includegraphics[width=\textwidth]{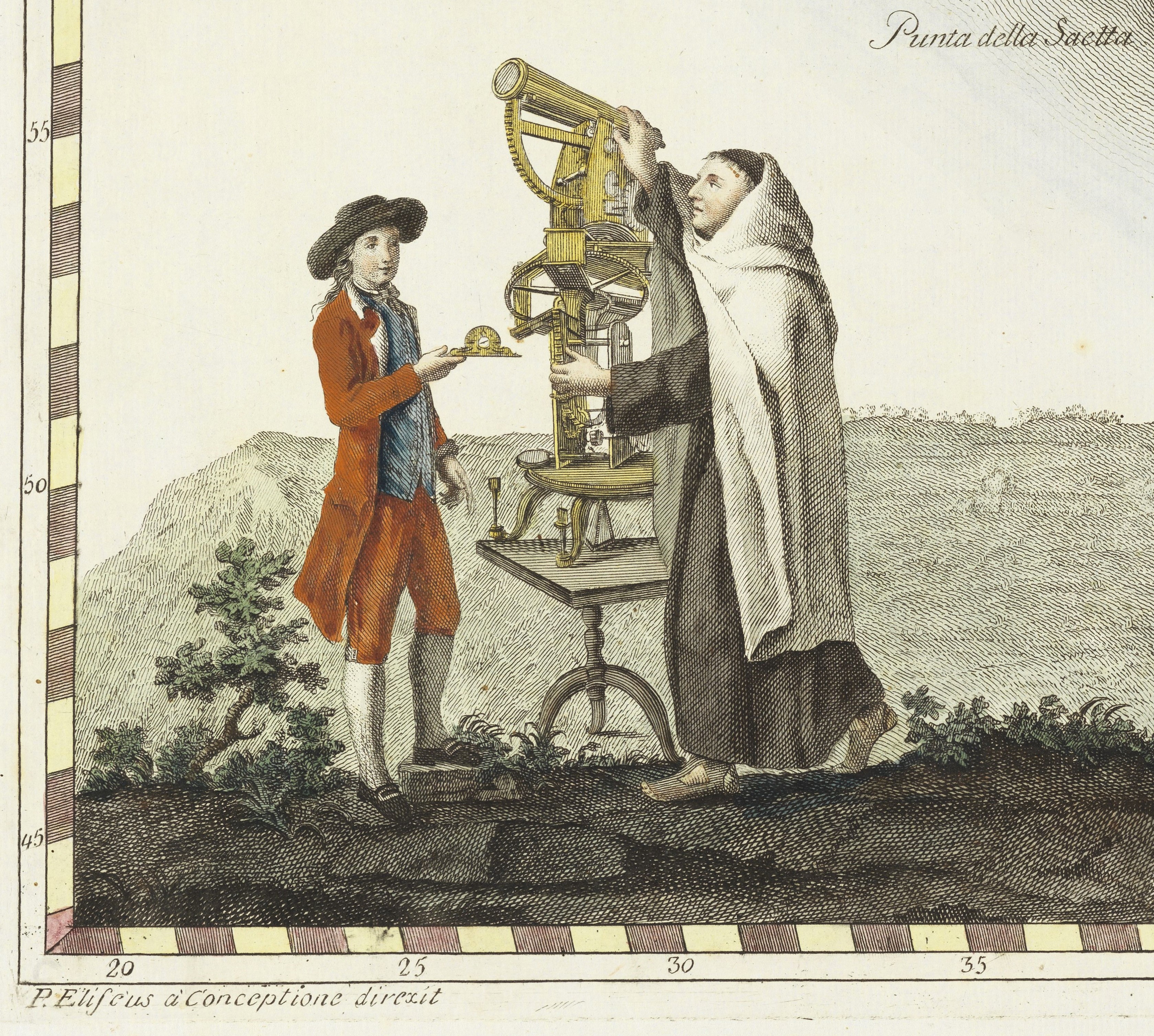}
\caption{Raffigurazione di p.~Eliseo con la sua macchina equatoriale, tratta dalla \emph{Carta corografica della Calabria ulteriore} del 1784, Bibliothèque nationale de France~\cite{Eliseo-Gallica}.}
\label{fig:Eliseo-Gallica}       			
\end{figure}

Per dare stabilità all'insegnamento della fisica, la Deputazione ricorse nuovamente a p.~Eliseo.
Adesso, grazie al cospicuo stipendio di 300 onze l'anno, p.~Eliseo non aveva difficoltà a trasferirsi a Palermo e il 13 febbraio 1786 occupò la cattedra di Fisica Sperimentale presso la Regia Accademia~\cite{Cancila2006,Nastasi1998}.
La Fig.~\ref{fig:Eliseo-Gallica} mostra una raffigurazione di p.~Eliseo con la sua macchina equatoriale, tratta dalla \emph{Carta corografica della Calabria ulteriore} del 1784, custodita presso la Bibliothèque nationale de France~\cite{Eliseo-Gallica}.

\FloatBarrier

Padre Eliseo portando con sé molti suoi strumenti incentivò lo studio della fisica e creò un nuovo e diverso interesse nei confronti di questa disciplina.
A tale proposito, Scinà che forse gli fu allievo e che gli successe nell'insegnamento scrive~\cite[p.~357]{Scina1859}:

\begin{quote}
``\emph{Tra i molti, che furono invitati, si ebbe il p.~Eliseo della Concezione da Napoli, che seco portò i suoi strumenti copiosi di numero, e se non esatti, sufficienti almeno a recare ad effetto tutti quegli esperimenti, che dai fisici sino allora erano stati immaginati.
Il perché i nostri dalla vista delle macchine e dell'esperienze corsero lieti alle lezioni del p.~Eliseo, e la fisica cominciò in Palermo a riguardarsi, come studio necessario alla cultura, ed all'avanzamento degl'ingegni nelle scienze.}''
\end{quote}

\noindent Oltre a incentivare lo studio della fisica con le sue ``macchine'', durante la sua permanenza a Palermo, p.~Eliseo pubblicò in latino due trattati di elementi di fisica sperimentale:

\begin{itemize}
  \item \emph{Physicae Experimentalis Elementa R. Panormitanae Academiae usui Et Experimentis publice instituendis accommodata auctore P. Eliseo a Conceptione carmelita excalceato. Volumen Unicum PHYSICAM GENERALEM COMPLECTENS. Panormi MDCCXC Typis Regiis}
  \item \emph{Physicae Experimentalis Elementa R. Panormitanae Academiae usui Et Experimentis publice instituendis accommodata auctore P. Eliseo a Conceptione carmelita excalceato PHYSICAE PARTICULARIS}
  \begin{itemize}
    \item \emph{PARS I Complectens. Aerostaticam, Aerologiam, \& Meteorologiam. Panormi MDCCXXXIX Typis Regiis}
    \item \emph{PARS II Thermologiam, Opticam, Dioptricam, \& Catoptricam. Panormi MDCCXC Typis Regiis}
    \item \emph{PARS III \& IV, UBI de Hydrostatica, Hydrodynamica, Hydraulica, Hydrologia \& de iis quae ad terram pertinent agitur. Panormi MDCCXC Typis Regiis}
  \end{itemize}
\end{itemize}

\noindent I trattati furono stampati nel 1789/90 in più volumi nella nuova Reale Stamperia di Palermo, istituita dalla Deputazione con un dispaccio del 17 luglio 1779~\cite{Lentini2017,web2}:

\begin{quote}
``\emph{...fornita di tutte sorti di caratteri fatti a bella posta venir da fuori per potersi in essa dar giornalmente alle stampe tutte le carte, ordinazioni e dispacci.}''
\end{quote}

\begin{figure}[h!]
\centering
\includegraphics[width=0.5\textwidth]{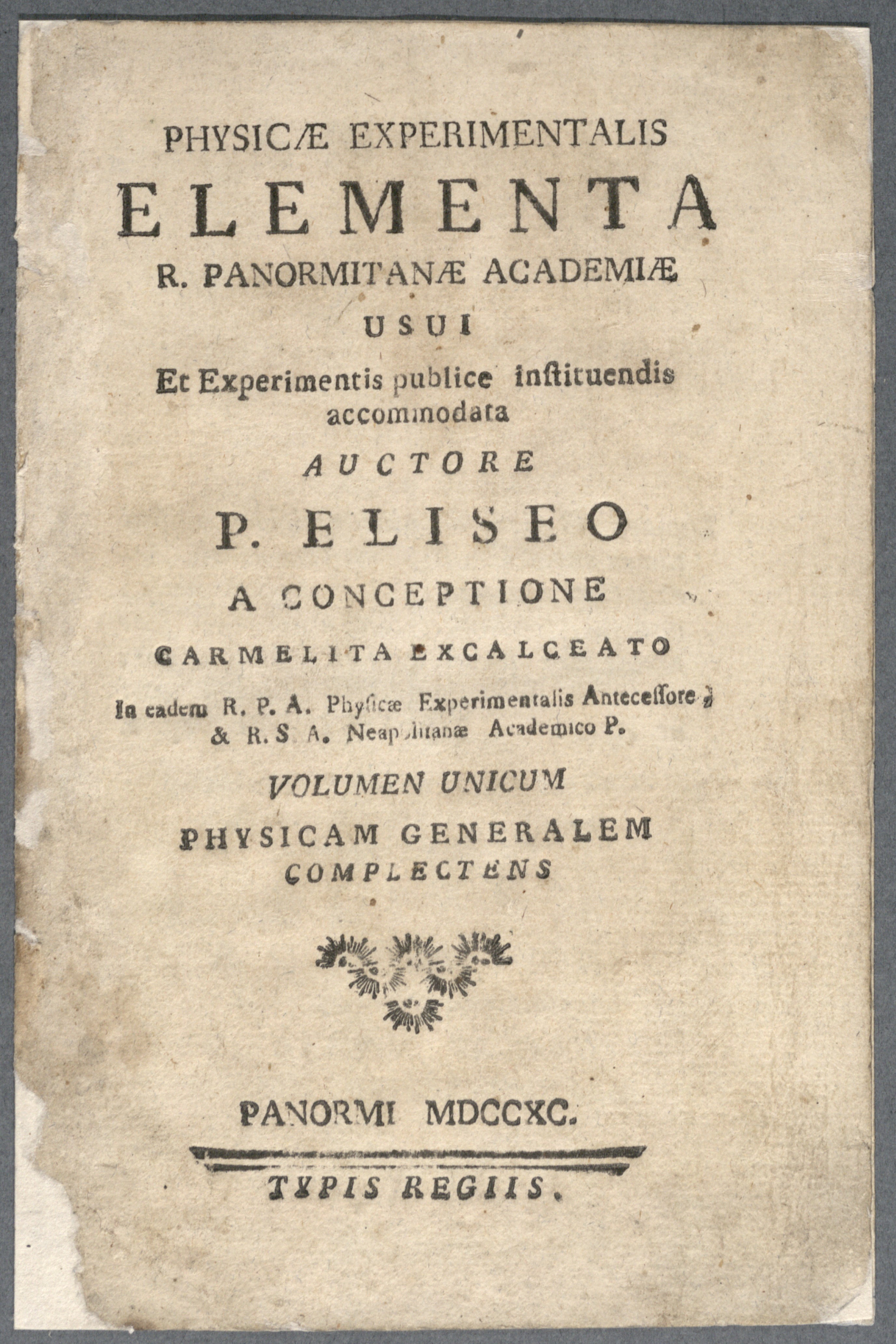}
\caption{Frontespizio del volume di p.~Eliseo della Concezione \emph{Physicae experimentalis elementa: physicam \mbox{generalem} complectens}.}
\label{fig:frontespizio}       			
\end{figure}

\begin{figure}[h!]
\centering \medskip
\includegraphics[width=0.8\textwidth]{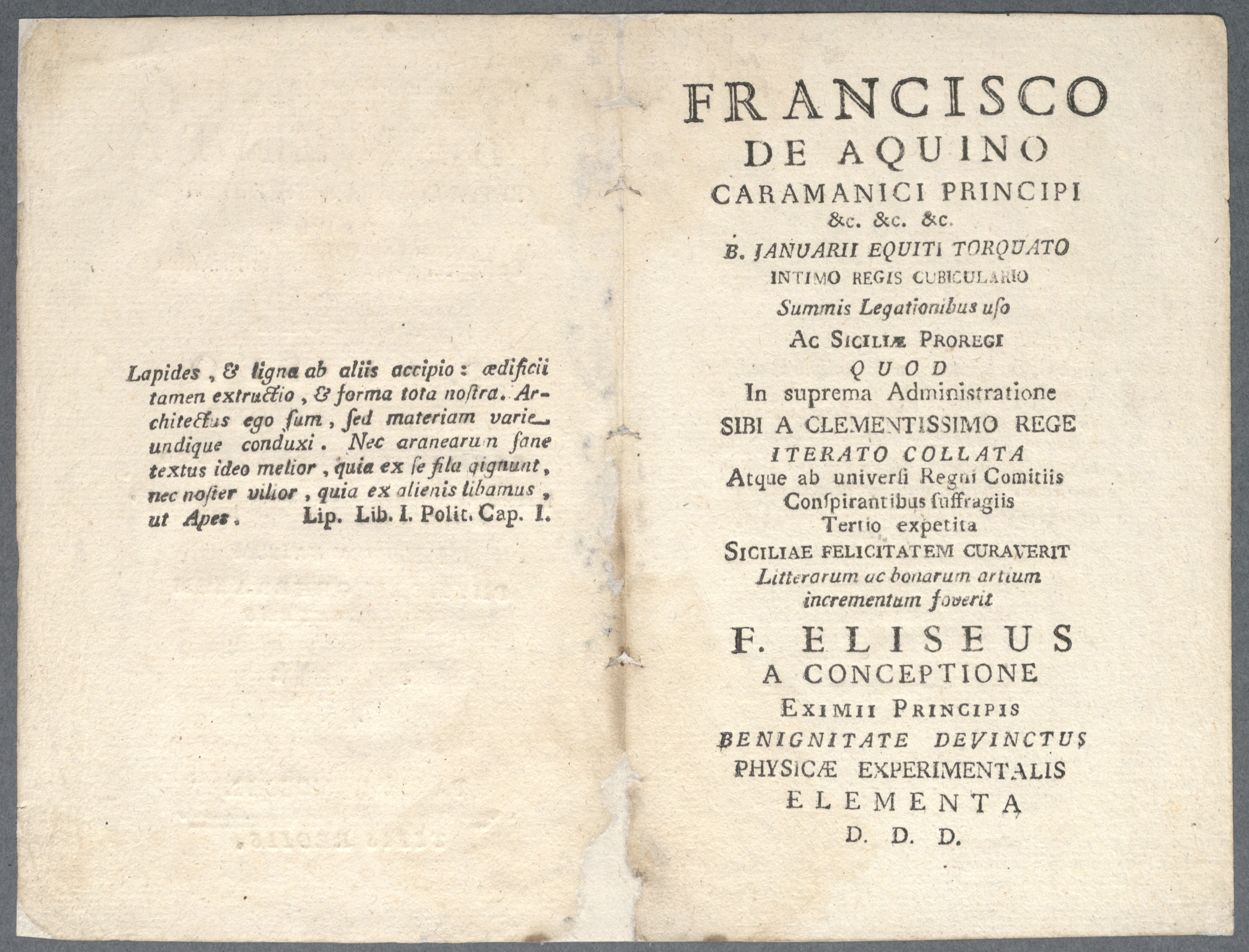}
\caption{Dedica al viceré di Sicilia Francesco D'Aquino principe di Caramanico (1738~-~1795).}
\label{fig:dedica}       			
\end{figure}

\noindent Nella Fig.~\ref{fig:frontespizio} è mostrato il frontespizio del volume \emph{Physicae experimentalis elementa: physicam generalem complectens} e nella Fig.~\ref{fig:dedica} la dedica al viceré di Sicilia Francesco D'Aquino principe di Caramanico (1738~-~1795).

\vfill

\vfill

L'istituzione della Reale stamperia consentì, quindi, la circolazione su carta stampata di opere straniere, tradotte dall'inglese, dal tedesco o dal francese, contribuendo in modo fondamentale alla circolazione delle idee che si andavano affermando in Europa.
Essa ebbe dunque una parte importante per la creazione di una Università che consentisse l'accesso all'istruzione di grado superiore anche ai ceti meno abbienti.

Padre Eliseo si dedicò costantemente alla ricerca scientifica nel campo della fisica, della chimica pneumatica e degli aeriformi~\cite{Zingales2022}; adottando il metodo sperimentale di Newton, analizzò tutti i campi di ricerca della fisica dell'epoca e corredò ogni sezione di un'ampia prospettiva storica sulle diverse teorie e di una serie di esperimenti pratici, illustrati nelle relative Tavole presenti nei due trattati.
Tuttavia, il manuale di p.~Eliseo, inadeguato rispetto alle nuove teorie che si andavano sviluppando nel campo della fisica e della chimica, non ebbe una grande diffusione.
Questi \emph{Elementi} di p.~Eliseo, scrive Scinà, destinati a sopperire alle manchevolezze dell'ormai superato manuale di Muscembroechio\footnote{Petrus van Musschenbroek, professore di astronomia a Leida, era noto per aver ideato e descritto per la prima volta nel 1731 uno strumento per lo studio della dilatazione termica dei metalli a cui diede anche il nome di \emph{pirometro}, oggi noto col nome di \emph{dilatometro termico lineare}, per i suoi esperimenti di elettrostatica del 1746 con i primordiali condensatori elettrostatici, noti con il nome di \emph{bottiglia di Leida}, e per i suoi trattati di fisica.} (Petrus van Musschenbroek, 1692~-~1761)~\cite{Esposito2014} e del manuale di George Atwood\footnote{George Atwood era noto per avere ideato nel 1776 -- 1779 una macchina per lo studio della dinamica, nota con il nome di \emph{macchina di Atwood}, che consente di illustrare agevolmente la legge del moto uniformemente accelerato.} (1745~-~1807)~\cite{Esposito2014}, quest'ultimo introdotto a Palermo nel 1781, erano però fondati ``\emph{sopra vecchie e cadenti opinioni, appena nati perirono}''~\cite[p.~359]{Scina1859}.

Nel 1793, ottenuto un congedo per motivi di salute, p.~Eliseo se ne tornava a Napoli ``appaltando'' per 150 onze l'anno l'incarico delle lezioni al già citato Nicolò Fresco.
Pare però che il Fresco, avendo a Napoli altri incarichi, abbia preferito ``subappaltare'' a Scinà l'incarico dell'insegnamento della Fisica Sperimentale con lo stipendio di sole 40 onze annue.
Una situazione deprimente per l'Accademia palermitana e umiliante per Scinà che, senza alternative per quella professione di docente universitario, dovette subirla fino al 1811 quando, in seguito alla morte di p.~Eliseo, il 7 gennaio 1809, poté godere la tanto sospirata ``proprietà'' della cattedra di Fisica Sperimentale~\cite{Nastasi1998}.

\section{Discussione}
\begin{figure}[h!]
\centering
\includegraphics[trim={0 25mm 0 8mm},clip=true,width=\textwidth]{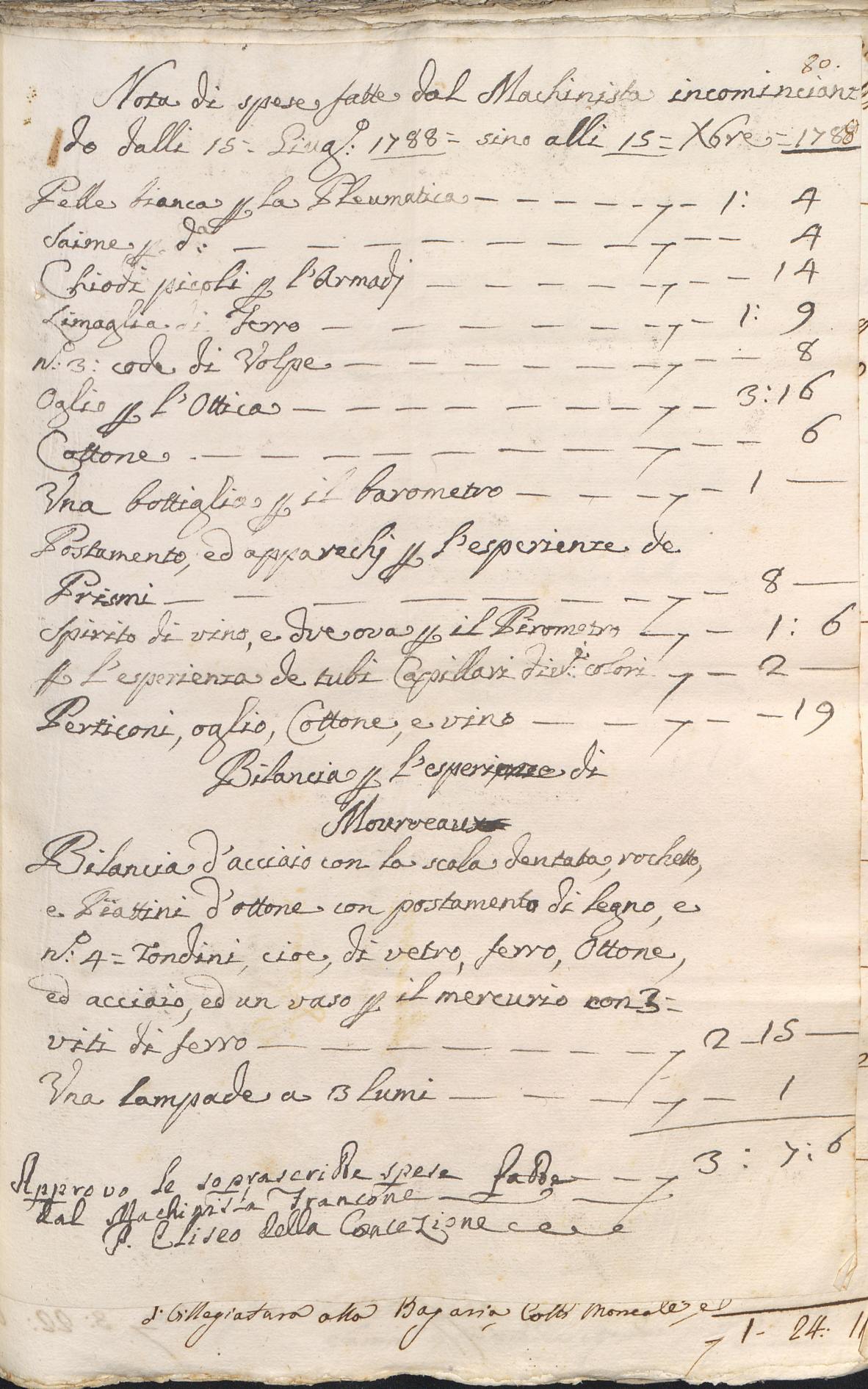}
\caption{Nota di spesa del 1788 effettuata dal macchinista Giovanni Francone a firma di p.~Eliseo della Concezione, conservata presso l'Archivio Storico dell'Università di Palermo~\cite{cautela622}.}
\label{fig:nota-spesa}       			
\end{figure}

In occasione delle celebrazioni del 300° anniversario della nascita di p.~Eliseo della Concezione, sono state condotte ricerche storiche riguardanti l'attività svolta presso la Regia Accademia di Palermo.
Grazie alla collaborazione del personale dello SBA, sono state individuate una serie di cautele riguardanti il periodo storico in cui p.~Eliseo operò presso l'Accademia palermitana.
I documenti rinvenuti riguardano principalmente note di spesa per materiale di consumo e strumenti scientifici.
La Fig.~\ref{fig:nota-spesa} mostra una nota del 1788 effettuata dal macchinista Giovanni Francone a firma di p.~Eliseo della Concezione, conservata presso l'Archivio Storico dell'Università di Palermo~\cite{cautela622}.

\begin{figure}[h!]
\centering
\includegraphics[width=0.9\textwidth]{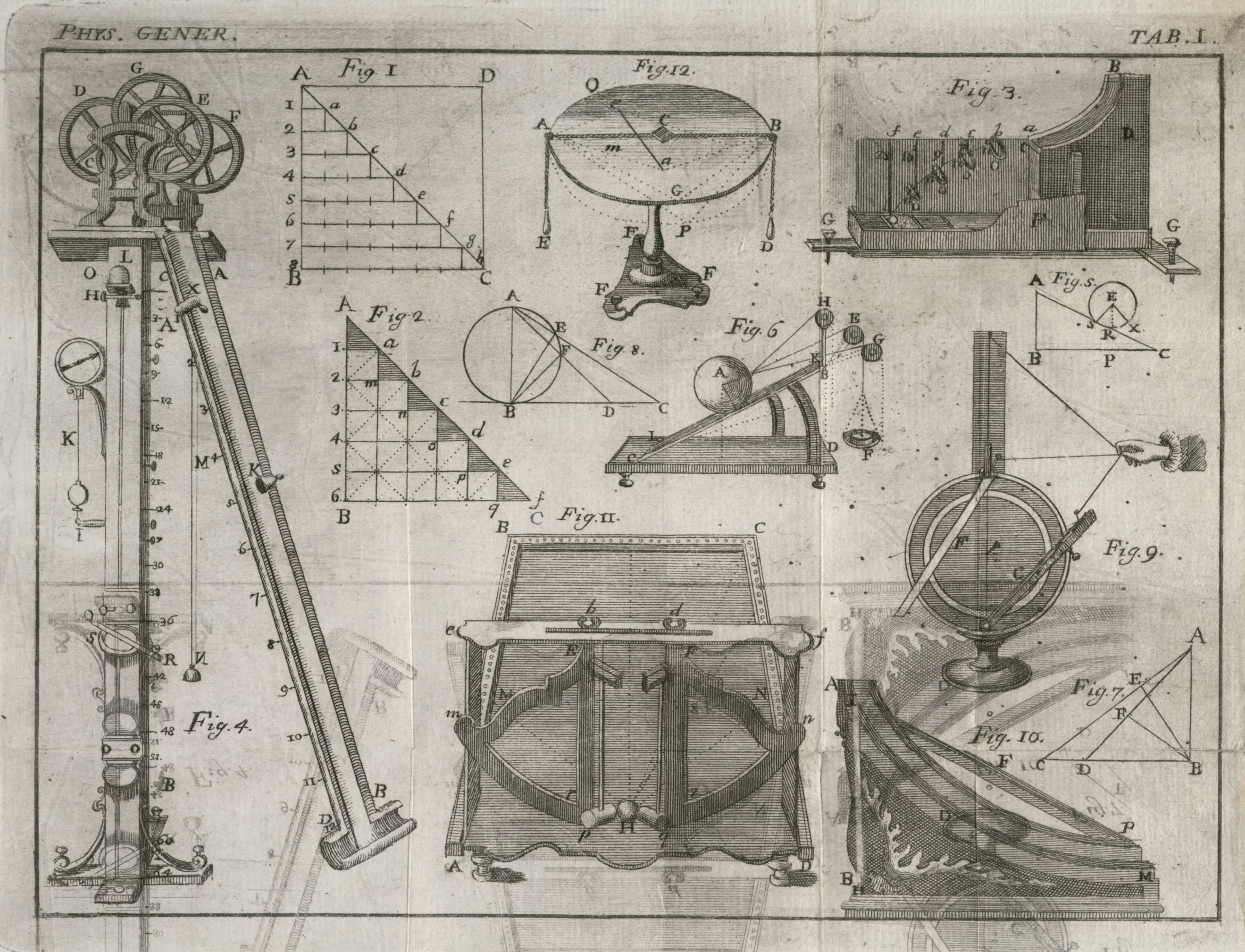}
\caption{Tavola I, \emph{Physicae experimentalis elementa: physicam generalem complectens}.}
\label{fig:Tavola1}       			
\end{figure}

\begin{figure}[h!]
\centering
\includegraphics[width=0.9\textwidth]{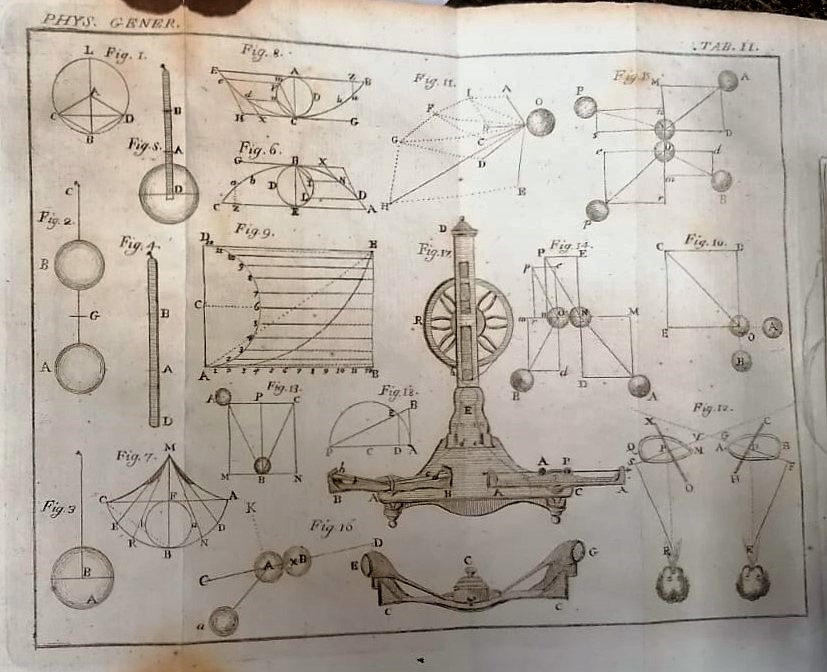}
\caption{Tavola II, \emph{Physicae experimentalis elementa: physicam generalem complectens}.}
\label{fig:Tavola2}       			
\end{figure}

\begin{figure}[h!]
\centering
\includegraphics[width=0.9\textwidth]{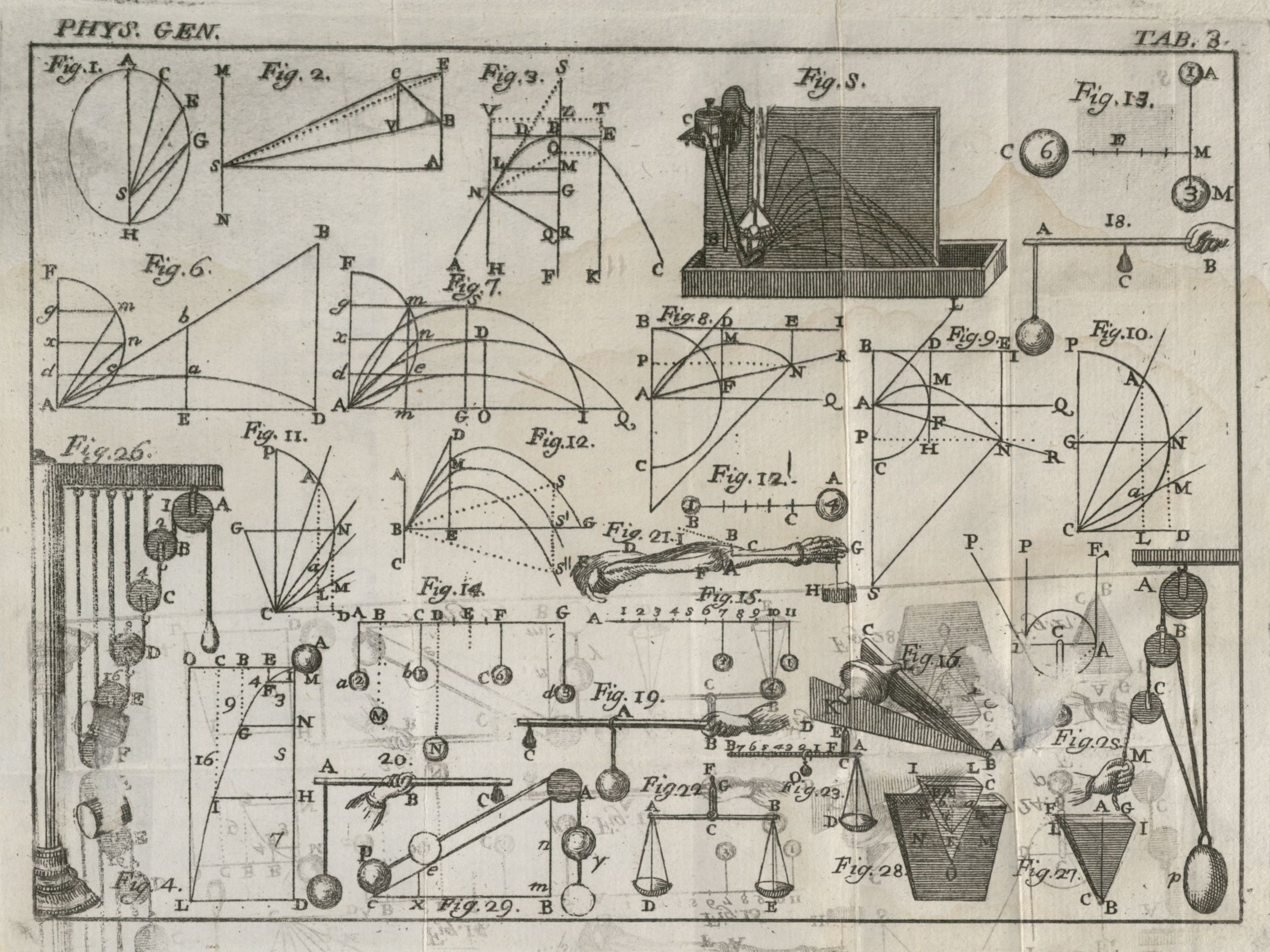}
\caption{Tavola III, \emph{Physicae experimentalis elementa: physicam generalem complectens}.}
\label{fig:Tavola3}       			
\end{figure}

\begin{figure}[h!]
\centering \bigskip
\includegraphics[width=0.9\textwidth]{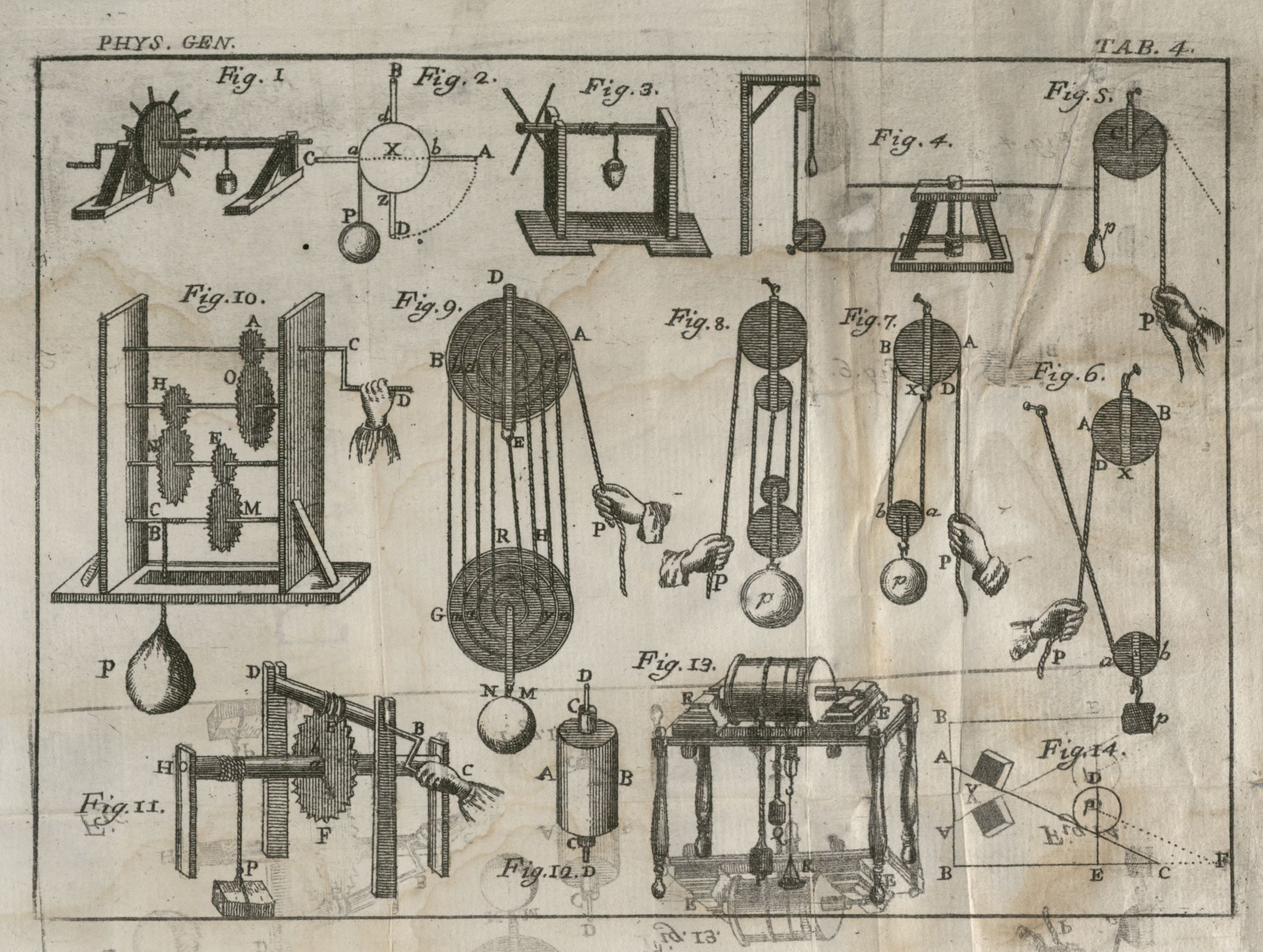}
\caption{Tavola IV, \emph{Physicae experimentalis elementa: physicam generalem complectens}.}
\label{fig:Tavola4}       			
\end{figure}

Recentemente, un esemplare del volume di p.~Eliseo, \emph{Physicae Experimentalis elementa: physicam generalem complectens}, acquistato sul libero mercato, è stato restaurato.
Gli interventi sono stati effettuati nell'ambito delle attività laboratoriali del Corso di Laurea Magistrale a ciclo unico in \emph{Conservazione e Restauro dei Beni Culturali} dell'Università di Palermo, seguendo i principi della riconoscibilità e reversibilità e della compatibilità dei materiali impiegati~\cite{Ansaldo2024}.
Il trattato è corredato da quattro tavole con incisioni calcografiche che riproducono schemi descrittivi delle leggi fisiche e illustrano strumenti di meccanica.
La Tavola II era mancante.
Grazie al confronto con un altro esemplare, custodito nella biblioteca storica dell'Osservatorio Astronomico ``Giuseppe S.~Vaiana'' di Palermo, la tavola mancante è stata identificata.
Le quattro tavole sono riportate rispettivamente nelle Figg.~\ref{fig:Tavola1},~\ref{fig:Tavola2},~\ref{fig:Tavola3}~e~\ref{fig:Tavola4}.
Infine, per consentire una più facile futura fruizione e quindi una maggiore diffusione del contributo scientifico di p.~Eliseo, anche da parte di un ampio pubblico, in fase di restauro il volume è stato digitalizzato.

Prendendo spunto proprio dalle tavole, in particolare dalla Tavola III (Fig.~\ref{fig:Tavola3}) dove è illustrato il doppio cono che sale sul piano inclinato, è stato organizzato il laboratorio didattico ``Hands-on Science'' dal titolo ``Paradossi meccanici'', rivolto principalmente agli studenti di scuola superiore di primo e di secondo grado.

\FloatBarrier

\begin{figure}[h!]
\centering
\includegraphics[width=0.8\textwidth]{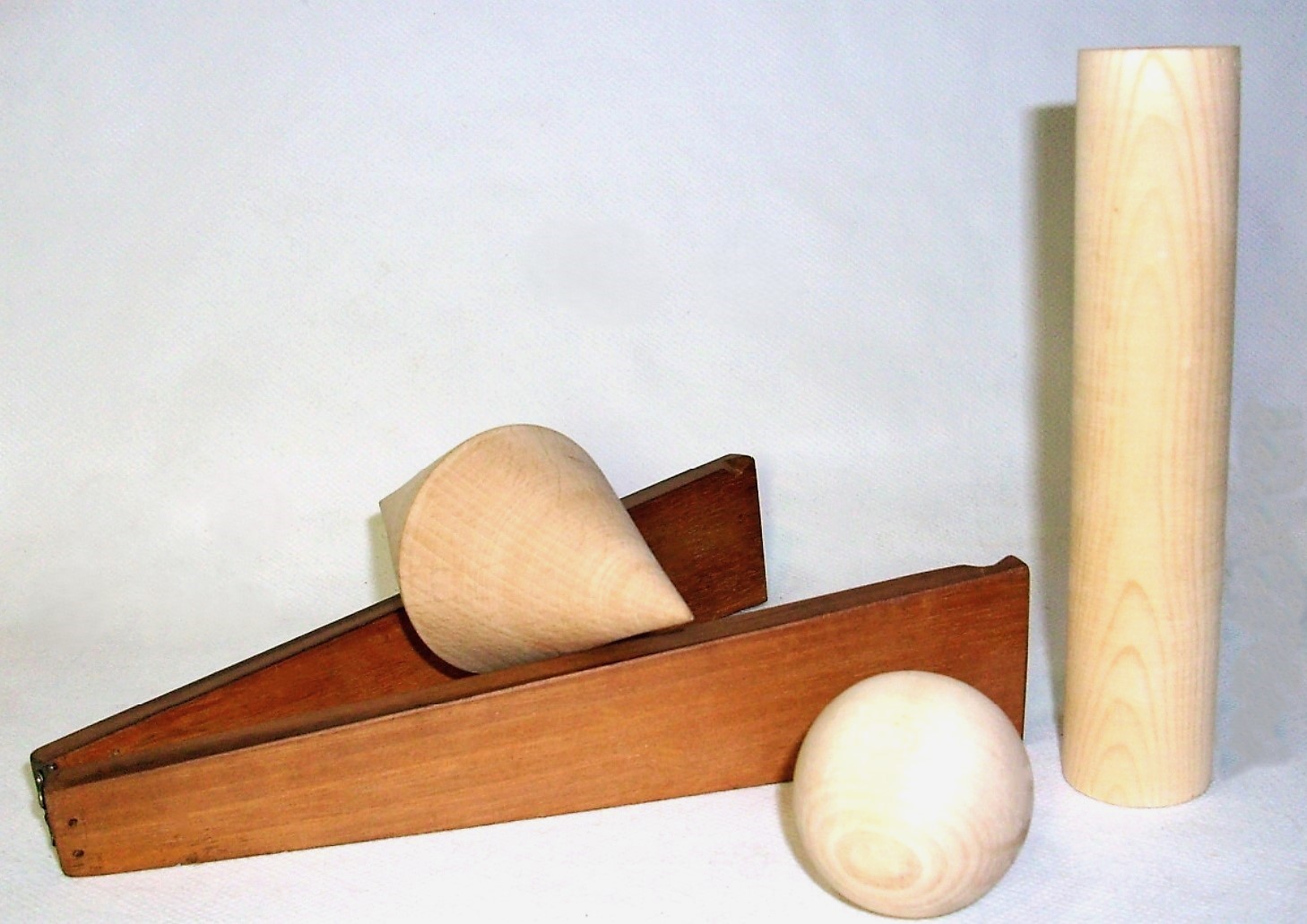}
\caption{Il doppio cono che sale sul piano inclinato; il cilindro e la sfera completano l'apparato.}
\label{fig:doppio-cono}       			
\end{figure}

\begin{figure}[h!]
\centering \bigskip
\includegraphics[width=0.8\textwidth]{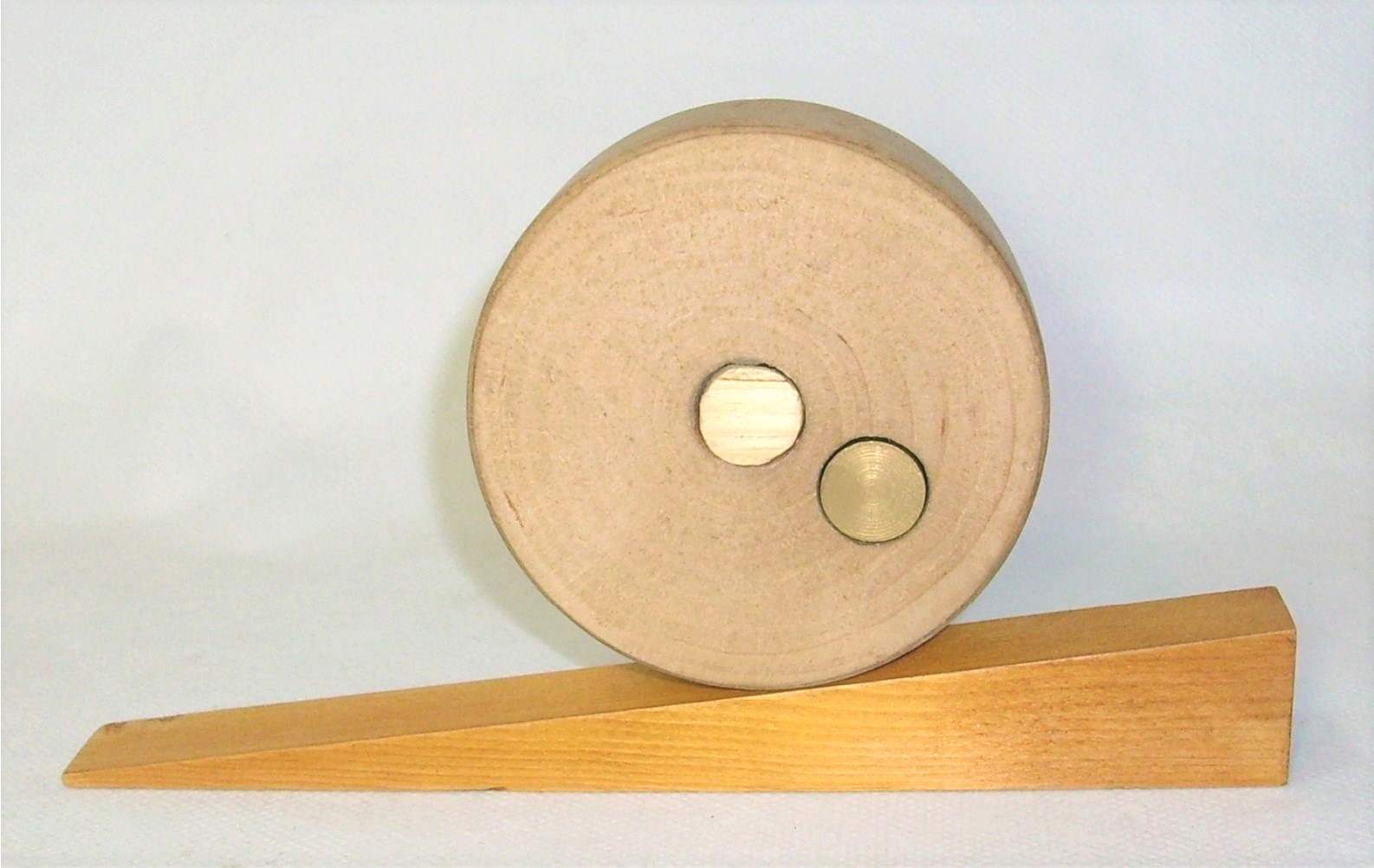}
\caption{Il cilindro con zavorra in equilibrio sul piano inclinato.}
\label{fig:cilindro}       			
\end{figure}

\begin{figure}[h!]
\centering \bigskip
\includegraphics[width=0.5\textwidth]{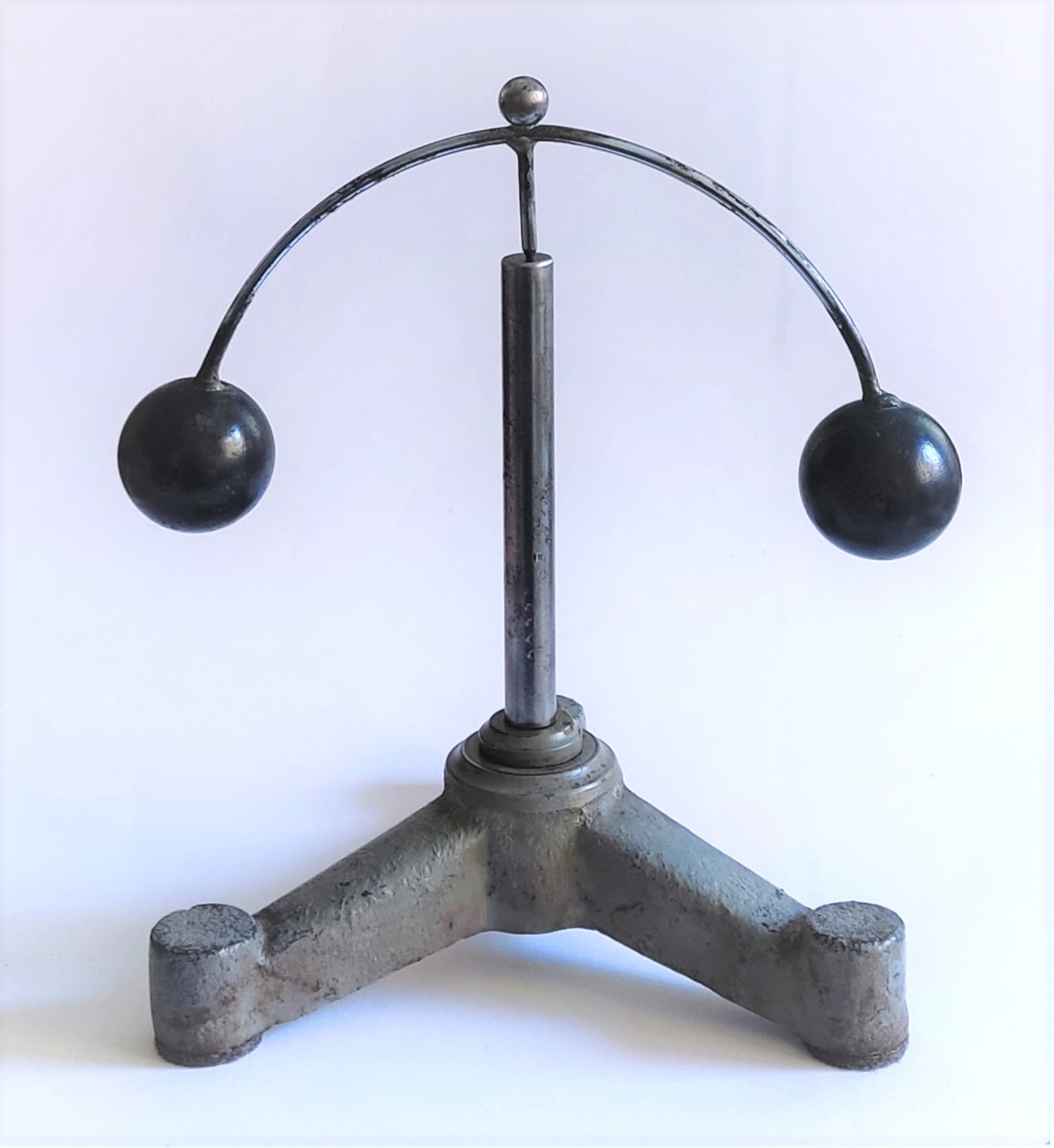}
\caption{Apparato dell'equilibrista per illustrare la condizione di equilibrio stabile.}
\label{fig:equilibrista}       			
\end{figure}

\begin{figure}[h!]
\centering\smallskip
\includegraphics[width=0.85\textwidth]{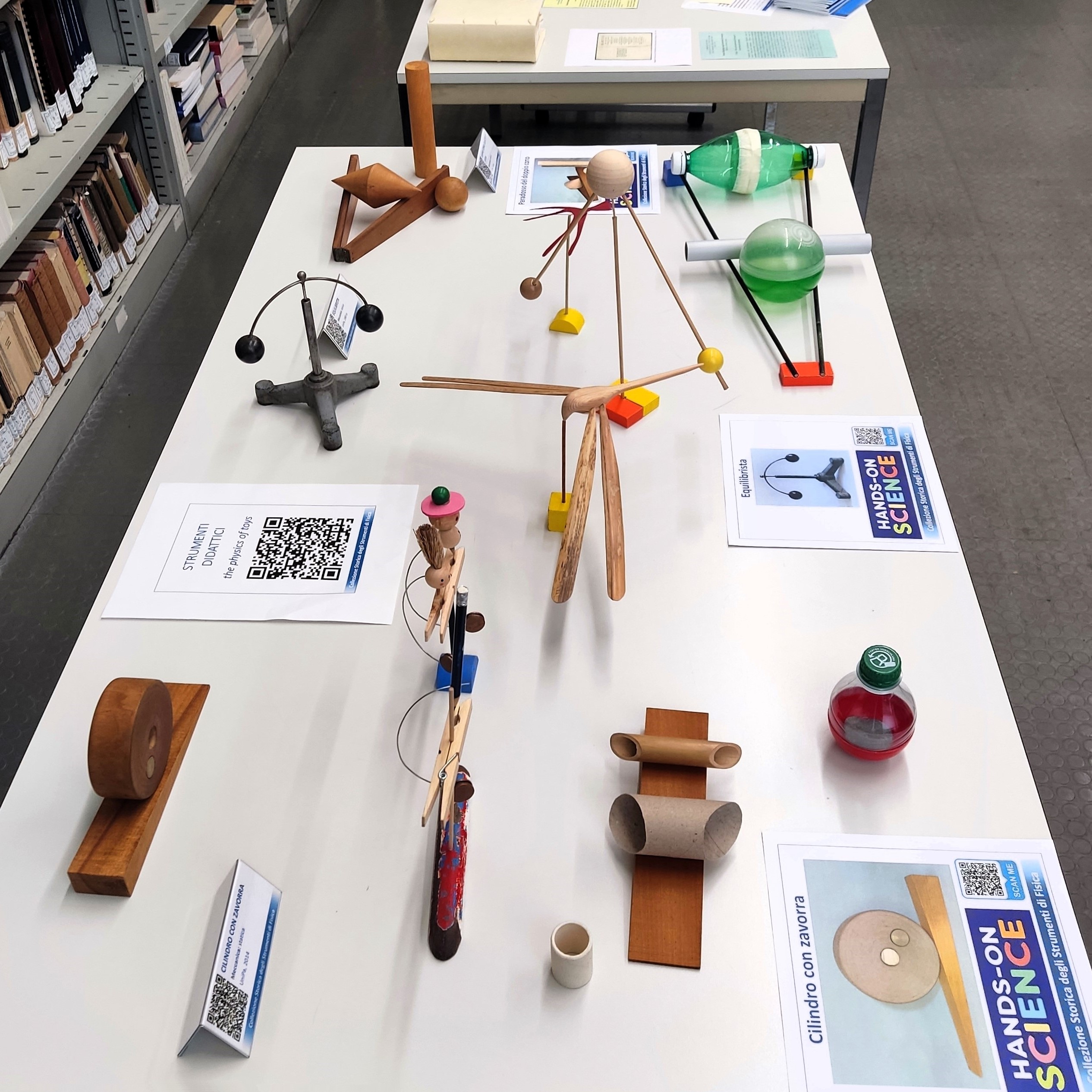}
\caption{Allestimento espositivo.}
\label{fig:allestimento}       			
\end{figure}

Sono stati predisposti tre differenti apparati didattici: il doppio cono (Fig.~\ref{fig:doppio-cono}), il cilindro impiombato (Fig.~\ref{fig:cilindro}) e l'equilibrista (Fig.~\ref{fig:equilibrista}), corredati da schede descrittive.
Questi apparati, mostrati nella Fig.~\ref{fig:allestimento}, permettono di eseguire esperimenti sorprendenti che contraddicono il senso comune e per tale motivo vengono denominati ``paradossi meccanici''.
Tuttavia, dopo un'attenta osservazione e un'analisi più approfondita, si può constatare che i fenomeni sono perfettamente coerenti con le leggi della fisica~\cite{Agliolo2011,Aglione2013}.

\vfill

\vfill

\vfill

\vfill

\FloatBarrier

Quando il doppio cono sale sul piano inclinato, si ha l'impressione che esso violi le leggi di Newton, ma con un'attenta osservazione si può notare che il centro di massa (CM) del doppio cono si sposta verso il basso, rispettando così le leggi della meccanica~\cite{Leybourn1694,Desaguliers1745,Agliolo2011}.
Il cilindro impiombato nasconde una zavorra di ottone, inserita vicino al bordo, che sposta il CM del sistema lontano dall'asse del cilindro.
La zavorra crea quindi un momento meccanico che consente al cilindro di stare in equilibrio sopra un piano inclinato in una ben determinata posizione in cui tutti i momenti meccanici agenti sul cilindro si annullano~\cite{Desaguliers1745,Aglione2013}.
Infine, l'equilibrista consente di illustrare la condizione di equilibrio stabile di un corpo rigido.
Infatti grazie al manubrio con i pesi alle estremità, il CM del sistema si trova al di sotto del punto di appoggio sul sostegno.
Il manubrio rimane così in una posizione di equilibrio stabile~\cite{Desaguliers1745}.
Pertanto, l'uso di questi apparati consente agli studenti di sperimentare le proprietà del CM e dell'equilibrio stabile.

I paradossi meccanici possono essere utilizzati in generale per aumentare l'attenzione degli studenti durante le lezioni in classe e contribuire così ad accrescere l'interesse dei giovani per lo studio della fisica.
Vale la pena inoltre notare che sui principi fisici appena descritti si basano molti giocattoli meccanici, popolari in passato e molto spesso riproposti ancora oggi.

Gli strumenti illustrati nelle tavole del volume di p.~Eliseo (Figg.~\ref{fig:Tavola1},~\ref{fig:Tavola2},~\ref{fig:Tavola3}~e~\ref{fig:Tavola4}), così come le note di spesa (Fig.~\ref{fig:nota-spesa}), possono fornire importanti informazioni sull'origine degli strumenti più antichi della Collezione Storica degli Strumenti di Fisica, custodita nell'edificio storico di via Archirafi 36 del DiFC.
Essi infatti rappresentano una valida testimonianza e un importante mezzo per la ricostruzione dello sviluppo della fisica nei suoi aspetti storici e didattici, sin dalla fondazione della Regia Accademia, alla fine del XVIII secolo, grazie anche al lavoro svolto da p.~Eliseo durante la sua permanenza a Palermo~\cite{Sears2017,Agliolo2018,web3}.

Le collezioni scientifiche custodite nelle Università hanno un importante ruolo, che è quello di preservare la storia e il patrimonio culturale legato allo sviluppo delle specifiche discipline.
Oltre al fatto che l'approccio allo studio delle leggi fisiche basato sullo studio degli strumenti storici possa essere rilevante anche da un punto di vista didattico, per rafforzare la comprensione delle leggi fisiche e guidare gli studenti verso una completa e coerente conoscenza scientifica.
Pertanto, i musei e le collezioni scientifiche possono svolgere un ruolo centrale nel migliorare l'apprendimento delle materie scientifiche e in particolare della fisica~\cite{Pantano2010,Heering2015,Agliolo2017,Mujtaba2018,Agliolo2021}.

\section{Conclusione}
Nell'articolo sono state presentate le attività organizzate dalla Biblioteca di Fisica e Chimica dello SBA e il DiFC in occasione del 300° anniversario della nascita di padre Eliseo della Concezione, docente di Fisica Sperimentale nella Regia Accademia degli Studi di Palermo alla fine del XVIII secolo.
Sono state organizzate un'esposizione bibliografica e archivistica, visite guidate alla Collezione Storica degli Strumenti di Fisica e un laboratorio didattico ``Hands-on Science'' sui paradossi meccanici (il doppio cono, il cilindro impiombato e l'equilibrista).
Le scuole del territorio sono state invitate a partecipare, dando la possibilità agli studenti, sotto la guida dei loro insegnanti, di sperimentare direttamente con gli exhibit didattici allestiti per l'occasione.
Nell'articolo inoltre, dopo una breve descrizione biografica di p.~Eliseo, sono stati discussi gli aspetti storico-didattici del lavoro svolto da p.~Eliseo presso l'Accademia palermitana.

\section*{Ringraziamenti}
L'autore desidera ringraziare Angelika Ansaldo Patti, Claudio De Benedictis e Marco Di Bella per gli interventi di restauro del volume di p.~Eliseo e per i preziosi commenti e suggerimenti; Margherita Cinà, Germana Mulè, Marta Rubino e Francesca Tignola per la preziosa assistenza nella ricerca delle fonti bibliografiche storiche; Fabio Panfili per l'attenta lettura del manoscritto.

\begin{flushleft}

\end{flushleft}

\end{document}